\documentclass{svjour2}

\usepackage{euscript,amssymb,amsmath,graphicx,epsfig}

\newcommand{\be}[1]{\begin{equation}\label{#1}}
\newcommand{\ee}{\end{equation}}
\newcommand{\ba}[1]{\begin{eqnarray}\label{#1}}
\newcommand{\ea}{\end{eqnarray}}
\newcommand{\rf}[1]{(\ref{#1})}
\newcommand{\nn}{\nonumber}

\begin{document}

\title{Emergent quantum Euler equation\\ and Bose-Einstein condensates}

\author{Maxim V. Eingorn \and Vitaliy D. Rusov}

\institute{M. Eingorn \and V. Rusov \at Department of Theoretical and Experimental Nuclear Physics,\\ Odessa National Polytechnic University,\\
Shevchenko av. 1, Odessa 65044, Ukraine\\ \\
M. Eingorn \at Astronomical Observatory, Odessa National University,\\ Dvoryanskaya st. 2, Odessa 65082, Ukraine\\ \\
M. Eingorn \at Physics Department, North Carolina Central University,\\ Fayetteville st. 1801, Durham, North Carolina 27707, USA\\ \\
\email{maxim.eingorn@gmail.com}\\
\email{siiis@te.net.ua}\\}

\date{Received: date / Accepted: date}

\maketitle

\begin{abstract}
In this paper, proceeding from the recently developed way of deriving the quantum-mechanical equations from the classical ones, the complete system of
hydrodynamical equations, including the quantum Euler equation, is derived for a perfect fluid and an imperfect fluid with pairwise interaction between the
particles. For the Bose-Einstein condensate of the latter one the Bogolyubov spectrum of elementary excitations is easily reproduced in the acoustic
approximation.

\end{abstract}

\section*{Introduction}

\setcounter{equation}{0}

Scientific interest in the de Broglie-Bohm causal interpretation of quantum mechanics  \cite{Holland,Durr} and its applications has been recently appreciably
rekindled (see also, for example, \cite{Wyatt}). In particular, in \cite{Rusov}, following the ideas of N.G. Chetaev (the theorem on stable trajectories in
dynamics \cite{Chetaev}) and G.~'t~Hooft (the classical deterministic theory, supplemented with the mechanism of dissipation, generates the observed quantum
behaviour of our world \cite{Hooft}), it was explicitly demonstrated, that adding the dissipation energy $Q$ to the Hamilton function and imposing the
stability condition on motion of the corresponding mechanical system, one can come to the standard Schr\"odinger equation, as well as identify the famous
Bohmian potential exactly with the introduced dissipation energy. Besides, the indisputable advantage of the produced derivation of the Schr\"odinger equation
from Newtonian mechanics lies in the fact, that $|\psi|^2$ turns out to represent the density of trajectories in the configuration space.

One of the next natural steps lies in developing the hydrodynamical approach to the problem of motion and deriving the quantum Euler equation, and then
applying it for describing the Bose-Einstein condensates. This paper is devoted exactly to this very important and promising line of investigation.  Obviously,
if fundamental properties of quantum fluids are theoretically reproduced in the Bohmian mechanics in a natural way without any additional assumptions and
ungrounded extensions, than this fact may serve as one more circumstantial evidence of this theory.

\

The paper is organized in the following way. In Section 1 we derive the modified Liouville equation, taking into consideration quantum effects. Then we come to
the complete system of the quantum hydrodynamics equations in Section 2. Finally, in Section 3 we consider sound waves in a quantum imperfect fluid with
pairwise interaction between the particles and reproduce the Bogolyubov spectrum of elementary excitations for its Bose-Einstein condensate.  The main results
are summarized in Conclusion.

\

\section{Modified Liouville equation in de Broglie-Bohm theory}

One of the most logical and self-consistent methods for deriving the macroscopic hydrodynamical equations consists in averaging of the microscopic mechanical
equations of motion of each single particle. Following this prevalent in fluid physics method, let us consider a system of $N$ identical particles,  whose
motion obeys Newton's second law:
\be{1} \frac{d{\bf r}_i}{dt}={\bf v}_i,\quad m\frac{d{\bf v}_i}{dt}={\bf F}_i({\bf r}_k,t),\quad {\bf F}_i({\bf r}_k,t)=-\left.\frac{\partial U}{\partial {\bf
r}_i}\right|_t\, ,\ee
where ${\bf r}_i(t)$ and ${\bf v}_i(t)$ are the radius-vector and the velocity of the $i$-th particle respectively, $m$ is the mass  of a single particle,
${\bf F}_i$ is the force, acting on the $i$-th particle from the direction of the external field (for example, the gravitational one) and all other particles,
and $U({\bf r}_k,t)$ is the corresponding potential energy. For simplicity we restrict ourselves in this paper to the case of uncharged particles in the absent
external electromagnetic field\footnote{The more general case of charged particles in the present external electromagnetic field is a subject of our
forthcoming paper.}. Here and in what follows the index $k$ corresponds to the set of natural numbers from $1$ to $N$ inclusive.

In order to proceed from the microscopic equations \rf{1} to the macroscopic ones, one needs the $N$-particle distribution function $f_N({\bf r}_k,{\bf v}_k,t)$,
satisfying the standard normalization requirement
\be{2} \int f_N({\bf r}_k,{\bf v}_k,t)\prod\limits_{i=1}^Nd{\bf r}_id{\bf v}_i=1\ee
and the corresponding well-known continuity equation
\be{3} \frac{df_N}{dt}=\left.\frac{\partial f_N}{\partial t}\right|_{{\bf r}_k,{\bf v}_k}+\sum\limits_{i=1}^N{\bf v}_i\left.\frac{\partial f_N}{\partial {\bf
r}_i}\right|_{{\bf v}_k,t}+\frac{1}{m}\sum\limits_{i=1}^N{\bf F}_i({\bf r}_k,t)\left.\frac{\partial f_N}{\partial {\bf v}_i}\right|_{{\bf r}_k,t}=0\ee
in the phase space (here we treat ${\bf r}_k$ and ${\bf v}_k$ as independent variables). Along with \rf{2} the function $f_N({\bf r}_k,{\bf v}_k,t)$,
representing the probability density in the phase space, should satisfy the following evident equation:
\be{4} \int f_N({\bf r}_k,{\bf v}_k,t)\prod\limits_{i=1}^Nd{\bf v}_i=A^2({\bf r}_k,t)\, ,\ee
where the function $A({\bf r}_k,t)=|\psi({\bf r}_k,t)|$ is the modulus (amplitude) of the $N$-particle (total) wave  function $\psi({\bf r}_k,t)$, obeying the
Schr\"odinger equation\footnote{It should be noted, that there is no need to take into account the distribution of velocities, as distinct from the chapter 3.6
in \cite{Holland}.}
\be{5} i\hbar\left.\frac{\partial \psi}{\partial t}\right|_{{\bf r}_k}=-\frac{\hbar^2}{2m}\sum\limits_{i=1}^N\triangle_i\psi+U({\bf r}_k,t)\psi,\quad
\triangle_i=\left(\left.\frac{\partial}{\partial {\bf r}_i}\right|_{t}\right)^2\, .\ee

Obviously, as it directly follows from \rf{2} and \rf{4}, for the functions $A({\bf r}_k,t)$ and $\psi({\bf r}_k,t)$ the standard normalization requirement
\be{6} \int A^2({\bf r}_k,t)\prod\limits_{i=1}^Nd{\bf r}_i=\int |\psi({\bf r}_k,t)|^2\prod\limits_{i=1}^Nd{\bf r}_i=1\ee
holds true (naturally, $A^2({\bf r}_k,t)=|\psi({\bf r}_k,t)|^2$ represents the probability density in the configuration space).  Let us also note, that the
functions $f_N({\bf r}_k,{\bf v}_k,t)$ and $A({\bf r}_k,t)$ are invariant relative to permutation of two arbitrary particles in view of their identity.

According to the de Broglie-Bohm theory \cite{Holland,Durr,Wyatt,Rusov}, the potential energy $U({\bf r}_k,t)$ may be presented in the form
\be{7} U({\bf r}_k,t)=U_{\mathrm{cl}}({\bf r}_k,t)+Q({\bf r}_k,t)\, ,\ee
where $U_{\mathrm{cl}}({\bf r}_k,t)$ is the classical part, and $Q({\bf r}_k,t)$ is the Bohmian quantum potential:
\be{8} Q({\bf r}_k,t)=-\frac{\hbar^2}{2m}\sum\limits_{j=1}^N\frac{\triangle_jA}{A}\, .\ee

Expressing $A({\bf r}_k,t)$ from \rf{4} and substituting the result into \rf{8}, we obtain
\be{9} Q({\bf r}_k,t)=-\frac{\hbar^2}{2m}\sum\limits_{j=1}^N\frac{\triangle_j\sqrt{\int f_N({\bf r}_k,{\bf v}_k,t)\prod\limits_{l=1}^Nd{\bf v}_l}} {\sqrt{\int
f_N({\bf r}_k,{\bf v}_k,t)\prod\limits_{l=1}^Nd{\bf v}_l}}\, .\ee

Substituting \rf{1}, \rf{7} and \rf{9} into \rf{3}, we come to the modified Liouville equation (see also \cite{Sparber})
\ba{10} &{}&\left.\frac{\partial f_N}{\partial t}\right|_{{\bf r}_k,{\bf v}_k}+\sum\limits_{i=1}^N{\bf v}_i \left.\frac{\partial f_N}{\partial {\bf
r}_i}\right|_{{\bf v}_k,t}-\frac{1}{m}\sum\limits_{i=1}^N\left.\frac{\partial U_{\mathrm{cl}}}{\partial {\bf r}_i}\right|_t
\left.\frac{\partial f_N}{\partial {\bf v}_i}\right|_{{\bf r}_k,t}+\nn\\
&+& \frac{\hbar^2}{2m^2}\sum\limits_{i=1}^N\left.\frac{\partial}{\partial {\bf r}_i}\left(\sum\limits_{j=1}^N\frac{\triangle_j\sqrt{\int f_N({\bf r}_k, {\bf
v}_k,t)\prod\limits_{l=1}^Nd{\bf v}_l}}{\sqrt{\int f_N({\bf r}_k,{\bf v}_k,t)\prod\limits_{l=1}^Nd{\bf v}_l}}\right)\right|_t\left.\frac{\partial f_N}{\partial
{\bf v}_i}\right|_{{\bf r}_k,t}=0\, .\ea

Obviously, the modification lies in the last term of quantum nature, proportional to $\hbar^2$. In contrast to three previous terms, it is essentially
nonlinear concerning $f_N({\bf r}_k,{\bf v}_k,t)$. It should be also mentioned, that the obtained equation \rf{10} contains all information  about the system.
It other words, strictly speaking, it is not necessary to solve any other additional equation, for example, the Schr\"odinger equation \rf{5}, for solving
\rf{10}, where all quantum features are already taken into account.

On the other hand, the modified Liouville equation \rf{10} is very difficult to solve, so one needs to carry out some simplifications.  The clearest one of
them is considered in the next section.

\

\section{Equations of Bohmian hydrodynamics for a perfect fluid}

Let us give concrete expression to the classical potential energy $U_{\mathrm{cl}}({\bf r}_k,t)$, neglecting pairwise interaction between the particles  (or,
in other words, regarding our "fluid" as a perfect one):
\be{11} U_{\mathrm{cl}}({\bf r}_k,t)=\sum\limits_{j=1}^NV({\bf r}_j,t)\, ,\ee
where $V({\bf r}_j,t)$ is the potential energy of the $j$-th particle in the external field. For simplicity let us assume, that all $N$ particles are "in the same
quantum state", described by the normalized to unity wave function $\psi_0$ with the modulus $A_0$, then the solution of the Schr\"odinger equation \rf{5} may be
presented in the form
\be{12} \psi({\bf r}_k,t)=\prod\limits_{j=1}^N\psi_0({\bf r}_j,t),\quad i\hbar\frac{\partial \psi_0}{\partial t}=\left(-\frac{\hbar^2}{2m}\triangle+V({\bf
r},t)\right)\psi_0\, ,\ee
whence it follows, in particular, that
\be{13} A({\bf r}_k,t)=\prod\limits_{j=1}^NA_0({\bf r}_j,t)\, .\ee

It should be mentioned, that the function $A_0^2({\bf r},t)$ represents the macroscopic density of an arbitrary single particle.  Consequently, the macroscopic
density $n({\bf r},t)$ of the whole system is interconnected with $A_0^2({\bf r},t)$ as follows:
\be{14} n({\bf r},t)=NA_0^2({\bf r},t)\quad\Leftrightarrow\quad A_0({\bf r},t)=\sqrt{\frac{n({\bf r},t)}{N}}\, .\ee

Substituting \rf{13} and \rf{14} into \rf{8}, we get
\be{15} Q({\bf r}_k,t)=\sum\limits_{j=1}^NQ_1({\bf r}_j,t),\quad Q_1({\bf r}_j,t)=-\frac{\hbar^2}{2m}\frac{\triangle_jA_0({\bf r}_j,t)}{A_0({\bf
r}_j,t)}=-\frac{\hbar^2}{2m}\frac{\triangle_j\sqrt{n({\bf r}_j,t)}}{\sqrt{n({\bf r}_j,t)}}\, .\ee

Let us note, that the made assumption of the same quantum state for all $N$ particles enables to present the $N$-particle quantum potential $Q({\bf r}_k,t)$ in
the form of the sum of $N$ identical $1$-particle ones $Q_1({\bf r}_j,t)$, each depending on the radius-vector ${\bf r}_j$ of the corresponding $j$-th
particle, $j=1,2,\ldots,N$. Simultaneously the $N$-particle amplitude $A({\bf r}_k,t)$ is presented in the form of the product of $N$ identical $1$-particle
ones $A_0({\bf r}_j,t)$.  A similar presentation holds true for the $N$-particle distribution function:
\be{16} f_N({\bf r}_k,{\bf v}_k,t)=\prod\limits_{j=1}^Nf_1({\bf r}_j,{\bf v}_j,t)\, ,\ee
where the $1$-particle distribution function $f_1({\bf r},{\bf v},t)$ satisfies the normalization condition
\be{17} \int f_1({\bf r},{\bf v},t)d{\bf r}d{\bf v}=1\, .\ee

Thus, the same quantum state assumption is equivalent to the presentation \rf{16}, being generally used in fluid physics when the interparticle interaction may
be neglected. Taking into account more accurate presentations by means of correlation functions lies beyond the scope of the present paper.

As it directly follows from \rf{10} after the substitution of \rf{11} and \rf{16} (or from \rf{3} after the substitution of \rf{1}, \rf{7}, \rf{11} and
\rf{15}),  this function obeys the equation
\be{18} \left.\frac{\partial f_1}{\partial t}\right|_{{\bf r},{\bf v}}+{\bf v}\left.\frac{\partial f_1}{\partial {\bf r}}\right|_{{\bf
v},t}-\frac{1}{m}\left.\frac{\partial}{\partial {\bf r}}\left(V({\bf r},t)-\frac{\hbar^2}{2m}\frac{\triangle\sqrt{n({\bf r},t)}}{\sqrt{n({\bf
r},t)}}\right)\right|_t\left.\frac{\partial f_1}{\partial {\bf v}}\right|_{{\bf r},t}=0\, .\ee

Applying the standard methods of fluid physics (in particular, presenting ${\bf v}$ in the form of the sum of the hydrodynamical velocity field ${\bf u}({\bf
r},t)$ and the deviation $\delta{\bf v}$ from it), one can easily obtain from \rf{18} the complete system of hydrodynamical equations,  describing motion of
our perfect fluid:
\be{19} \frac{\partial \rho}{\partial t}+\frac{\partial}{\partial {\bf r}}\left(\rho{\bf u}\right)=0\, ,\ee
\be{20} \frac{\partial {\bf u}}{\partial t}+\left({\bf u}\frac{\partial}{\partial{\bf r}}\right){\bf u}=-\frac{1}{\rho}\frac{\partial p}{\partial {\bf
r}}-\frac{1}{m}\frac{\partial V}{\partial {\bf r}}+\frac{\hbar^2}{2m^2}\frac{\partial }{\partial {\bf r}}\left(\frac{\triangle\sqrt{\rho}}{\sqrt{\rho}}\right)
\, ,\ee
\be{21} \frac{3}{2}\frac{\rho\kappa}{m}\left[\frac{\partial T}{\partial t}+\frac{\partial}{\partial {\bf r}}(T{\bf u})\right]= -\frac{\partial {\bf
q}}{\partial {\bf r}}\, ,\ee
where the mass density $\rho({\bf r},t)$, the pressure $p({\bf r},t)$, the temperature $T({\bf r},t)$ and, finally, the heat flux  density ${\bf q}({\bf r},t)$
are the macroscopic quantities, given by the following equations respectively:
\be{22} \rho({\bf r},t)=mn({\bf r},t)=Nm\int f_1({\bf r},{\bf v},t)d{\bf v}\, ,\ee
\be{23} p({\bf r},t)=n({\bf r},t)\kappa T({\bf r},t)=\frac{1}{3}Nm\int \delta v^2f_1({\bf r},{\bf v},t)d{\bf v}\, ,\ee
\be{24} {\bf q}({\bf r},t)=\frac{1}{2}Nm\int \delta v^2\delta{\bf v}f_1({\bf r},{\bf v},t)d{\bf v}\, .\ee

Here $\kappa$ is the Boltzmann constant, and viscosity is neglected. The continuity equation \rf{19} and the heat equation \rf{21}  exactly coincide with the
corresponding equations of classical (non-quantum) hydrodynamics \cite{Landau6}. Only the equation \rf{20}, which represents the quantum Euler equation,
differs from its classical analogue in the presence of the last term in the right hand side, proportional, in particular, to the ratio $\hbar^2/m^2$. Besides,
it is absolutely obvious, that one should expect the greater absolute value of the quantum potential $Q$, the sharper the spatial change of the mass density
$\rho$ is. If the fluid is incompressible, that is $\rho$ is simply some constant, then $Q=0$, and the distinction between classical and quantum Euler
equations is eliminated. In the general case the additional quantum term in \rf{20} vanishes, if the function $\sqrt{\rho}$ satisfies the Helmholtz equation
$\triangle\sqrt{\rho}+\mathrm{const}\sqrt{\rho}=0$, where $\mathrm{const}$ is an arbitrary constant.

Let us estimate, for which typical velocities and distances in some concrete problem this term, being nonzero, plays a vital part.  Introducing typical
velocity $u_0$ and distance $r_0$, one can immediately compare it with the second term in the left hand side of the same equation. They become quantities of
the same order for $u_0r_0\sim\hbar/m$. In this simple relationship one can easily recognize the famous Heisenberg's uncertainty principle, as it should be.
For example, if it is a question of helium, then the product $u_0r_0$ should be a quantity of the order $10^{-8}\, m^2/s$.

Thus, in this section we have derived the equations \rf{19}, \rf{20} and \rf{21} of Bohmian hydrodynamics for a perfect fluid.  Concluding this section, let us
note, that in the framework of standard quantum mechanics the hydrodynamical equations were written down by Landau and Khalatnikov (see, for example,
\cite{Brush}), but they contain quantum-mechanical operators with proper commutation rules and are much more complicated.

\

\section{Sound waves in imperfect fluid with pairwise interaction}

According to \cite{Landau6,Landau9}, only for temperatures, close to absolute zero, quantum effects advance to the forefront in properties of fluids,  but
actually only helium remains liquid down to absolute zero, while all other fluids become solid before quantum. As regards helium, our basic assumption of the
same quantum state for all particles holds true for the Bose-Einstein condensate of its isotope $^4\mathrm{He}$, being a Bose fluid (see also
\cite{Kadomtsev,Fetter}). Now we are interested in developing quantum acoustics for such quantum fluids, being, generally speaking, imperfect owing to
interparticle interaction.

In order to take proper account of pairwise interaction between the particles, one can add in the right hand side of \rf{11} the term
\be{25} V_{PI}=\frac{1}{2}\sum\limits_{j,l=1; j\neq l}^{N}\Phi({\bf r}_j-{\bf r}_l)\, ,\ee
where $\Phi({\bf r}_j-{\bf r}_l)$ is the potential energy of interaction of $j$-th and $l$-th particles, and then apply the method of correlation functions
for solving \rf{10}\footnote{Development of this method for the modified Liouville equation in the de Broglie-Bohm theory is also a subject of our forthcoming
paper.}, but there is also another way. Even in spite of interparticle interaction, the form of the equations from \rf{13} to \rf{16} may be preserved as a
certain approximation, if the "$1$-particle" Schr\"odinger equation \rf{12} is modified in the following way:
\be{26} i\hbar\frac{\partial \psi_0}{\partial t}=\left(-\frac{\hbar^2}{2m}\triangle+V({\bf r},t)+V_{GP}({\bf r},t)\right)\psi_0\, ,\ee
where the additional potential energy
\be{27} V_{GP}({\bf r},t)=\frac{4\pi\hbar^2 a}{m}NA_0^2({\bf r},t)=\frac{4\pi\hbar^2 a}{m^2}\rho({\bf r},t)\, ,\ee
where, in its turn, $a$ is a typical length of boson-boson scattering. The nonlinear $1$-particle Schr\"odinger equation \rf{26}  is well-known as the
Gross-Pitaevskii equation \cite{Gross,Pitaevskii} (see also \cite{Kadomtsev,Fetter}). Taking into consideration the presence of \rf{27}, instead of \rf{20} we
get the quantum Euler equation in the form
\be{28} \frac{\partial {\bf u}}{\partial t}+\left({\bf u}\frac{\partial}{\partial{\bf r}}\right){\bf u}= -\frac{1}{m}\frac{\partial V}{\partial {\bf
r}}-\frac{4\pi\hbar^2 a}{m^3}\frac{\partial\rho}{\partial {\bf r}}+\frac{\hbar^2}{2m^2}\frac{\partial }{\partial {\bf
r}}\left(\frac{\triangle\sqrt{\rho}}{\sqrt{\rho}}\right)\, .\ee

At the same time the continuity equation \rf{19} remains unchanged.

Now let us turn to the case $V=0$ and apply in the equations \rf{19} and \rf{28} the standard acoustic approximation: up to the zero order of smallness
$\rho=\rho_0$, ${\bf u}=0$ (where $\rho_0$ is some constant, depending neither on time $t$, nor on spatial coordinates ${\bf r}$), and up to the first order of
smallness $\rho=\rho_0+\rho_1$, ${\bf u}={\bf u}_1$ (where the additional terms $\rho_1$ and ${\bf u}_1$ are some functions of $t$ and ${\bf r}$). Then from
\rf{19} and \rf{28} up to the first order of smallness we obtain respectively
\be{29} \frac{\partial \rho_1}{\partial t}+\rho_0\frac{\partial{\bf u}_1}{\partial {\bf r}}=0\, ,\ee
\be{30} \frac{\partial {\bf u}_1}{\partial t}=-\frac{4\pi\hbar^2 a}{m^3}\frac{\partial\rho_1}{\partial {\bf r}}+\frac{\hbar^2}{4m^2\rho_0}\frac{\partial
}{\partial {\bf r}}\left(\triangle\rho_1\right)\, .\ee

From \rf{29} and \rf{30} the sound propagation equation follows immediately:
\be{31} \frac{\partial^2 \rho_1}{\partial t^2}-\frac{4\pi\hbar^2 a\rho_0}{m^3}\triangle\rho_1+\frac{\hbar^2}{4m^2}\triangle\left(\triangle\rho_1\right)=0\,
.\ee

Let us establish the dispersion law for plane waves. For that let us look for the solution of \rf{31} in the form
\be{32} \rho_1({\bf r},t)\sim\exp[i({\bf k}{\bf r})-i\omega t]\, ,\ee
where ${\bf k}$ is a wave vector, and $\omega$ is a frequency. Substituting \rf{32} into \rf{31}, we get
\be{33} \omega=\sqrt{\frac{4\pi\hbar^2 a\rho_0}{m^3}k^2+\frac{\hbar^2}{4m^2}k^4}\, ,\ee
where $k=|{\bf k}|$. In the vicinity of absolute zero of temperature there is the nonzero factor $4\pi\hbar^2 a\rho_0/m^3$ in front of $k^2$ in \rf{33}, which
may be interpreted, as usual, as the square of the sound speed due to pairwise interaction. The established dispersion law \rf{33} exactly coincides with the
famous Bogolyubov spectrum of elementary excitations in a Bose fluid. Thus, this fundamental spectrum is derived in our quantum-hydrodynamical approach in a
sufficiently simple way.

\

\section*{Conclusion}

In this paper, proceeding from the recently developed way of deriving the quantum-mechanical equations from the classical ones  \cite{Rusov}, we have derived
successively the modified Liouville equation (10) and the complete system of hydrodynamical equations (namely, the continuity equation (19), the quantum Euler
equation (20) and the heat equation (21)) for a perfect fluid, formed by identical particles, being in the same quantum state. This basic assumption holds true
for Bose-Einstein condensates of quantum fluids at very low temperatures. Then we have taken into account the interparticle interaction by using the
Gross-Pitaevskii equation (26). Finally, using the acoustic approximation, we have easily reproduced the Bogolyubov spectrum of elementary excitations (see
(33)).

In future, being based on the direct relativistic generalization \cite{RusovRelativity} of \cite{Rusov},  we shall also generalize the developed here quantum
hydrodynamics to the relativistic case. This line of investigation promises to open up new intriguing possibilities for studying, in particular, black holes
and the whole Universe.

\

\section*{Acknowledgements}

M.V. Eingorn wants to thank Prof. V.L. Kulinskii for very useful discussions.

V.D. Rusov is sincerely grateful to participants of the seminar of the Akhiezer Institute for Theoretical Physics (KITP, Kharkov, Ukraine)  for the benevolent
atmosphere during the discussion and the fruitful comments.

The work of M.V. Eingorn was supported in part by NSF CREST award HRD-0833184 and NASA grant NNX09AV07A.

\

\end{document}